\documentclass[aps,twocolumn,showpacs,preprintnumbers,nofootinbib,prd,superscriptaddress,groupedaddress,10pt]{revtex4-1}

\makeatletter
\def\l@subsubsection#1#2{}
\def\l@subsubsubsection#1#2{}
\makeatother

\usepackage{graphicx,amssymb,amsmath,amsthm,amsfonts,epsfig,epsf}
\usepackage[usenames]{color}
\usepackage{epstopdf}
\definecolor{darkred}{rgb}{0.5,0,0}

\usepackage{array}
\usepackage{afterpage}
\usepackage{bm}
\usepackage{dcolumn}
\usepackage[latin1]{inputenc}
\usepackage{latexsym}
\usepackage{rotating}
\usepackage{longtable}

\usepackage{enumerate}
\usepackage{tensor,multirow}
\usepackage{url}
\usepackage[linktocpage]{hyperref}
\usepackage{float}
\usepackage{graphicx}

\def\be{\begin{equation}}
\def\ee{\end{equation}}
\newcommand{\beq}{\begin{eqnarray}}
\newcommand{\eeq}{\end{eqnarray}}

\def\ba{\begin{align}}
\def\ea{\end{align}}

\newcounter{mnotecount}[section]

\renewcommand{\themnotecount}{\thesection.\arabic{mnotecount}}

\newcommand{\mnote}[1]
{\protect{\stepcounter{mnotecount}}$^{\mbox{\footnotesize
$
\bullet$\themnotecount}}$ \marginpar{
\raggedright\tiny\em
$\!\!\!\!\!\!\,\bullet$\themnotecount: #1} }

\begin{document}

\title{Strong cosmic censorship in charged black-hole spacetimes: still subtle}

\author{
Vitor Cardoso$^{1,2}$,
Jo\~ao L. Costa$^{3,4}$,
Kyriakos Destounis$^{1}$,
Peter Hintz$^{5}$,
Aron Jansen$^{6}$
}
\affiliation{${^1}$ CENTRA, Departamento de F\'{\i}sica, Instituto Superior T\'ecnico -- IST, Universidade de Lisboa -- UL,
Avenida Rovisco Pais 1, 1049 Lisboa, Portugal}
\affiliation{${^2}$ Perimeter Institute for Theoretical Physics, 31 Caroline Street North Waterloo, Ontario N2L 2Y5, Canada}
\affiliation{$^{3}$ Departamento de Matem\'atica, ISCTE - Instituto Universit\'ario de Lisboa, Portugal}
\affiliation{$^{4}$ Center for Mathematical Analysis, Geometry and Dynamical Systems, Instituto Superior T\'ecnico -- IST, Universidade de Lisboa -- UL,
Avenida Rovisco Pais 1, 1049 Lisboa, Portugal}
\affiliation{$^5$ Department of Mathematics, Massachusetts Institute of Technology, Cambridge, CA 02139, USA}
\affiliation{$^6$ Institute for Theoretical Physics and Center for Extreme Matter and Emergent Phenomena,
Utrecht University, 3508 TD Utrecht, The Netherlands}
\begin{abstract}
It was recently shown that Strong Cosmic Censorship may be violated in highly charged black-hole spacetimes living in a universe with a positive cosmological constant.
Several follow-up works have since suggested that such result, while conceptually interesting, cannot be upheld in practice.
We focus here on the claim that the presence of charged massive scalars suffices to save Strong Cosmic Censorship.
To the contrary, we show that there still exists a finite region in parameter space where Strong Cosmic Censorship is expected to be violated. 
\end{abstract}

\maketitle
%

\noindent{\bf{\em I. Introduction.}}
We recently presented an indication that Strong Cosmic Censorship (SCC) might be violated for charged, near-extremal Reissner--Nordstr\"{o}m (RN) black holes (BHs) in a de Sitter (dS) spacetime~\cite{Cardoso:2017soq}. More precisely, for linear massless and neutral scalar perturbations of RNdS BHs, the basic quantity controlling the stability of the Cauchy horizon, and therefore the fate of SCC, is given by~\cite{Hintz:2015jkj,CostaFranzen}
\begin{equation}
\label{betaOld}
\beta \equiv -\text{Im}(\omega_0)/\kappa_-\;,
\end{equation}
where $\omega_0$ is the longest-lived non-zero quasinormal
mode (QNM) and $\kappa_-$ is the surface gravity of the Cauchy horizon (CH).
Moreover, the results in~\cite{Hintz:2016gwb,Hintz:2016KNdS,CGNS4} suggest that $\beta$ remains the essential quantity in the non-linear setting: the higher $\beta$, the more stable the CH. Concretely, the modern formulation of SCC\footnote{inextendibility of metric coefficients in $H^1_{\rm loc}$ and of Christoffel symbols in $L^2_{\rm loc}$ across the CH.} demands that
\begin{equation}\label{theoremHintz}
\text{SCC} \leftrightarrow  \beta < 1/2
\end{equation}
in order to guarantee the breakdown of the field equation at the CH. One should also recall that $\beta<1$ is related to the blow up of curvature invariants.
In~\cite{Cardoso:2017soq}, a thorough numerical study of $\beta$ for the full range of subextremal RNdS spacetimes revealed, quite surprisingly, that $\beta>1/2$ in the near-extremal regime. However, it turned out that $\beta\leq 1$ always, with equality at extremal charge.
This provides evidence for the existence of Cauchy horizons which, upon perturbation, are rather singular due to the divergence of curvature invariants, but where the gravitational field can still be described by the field equations; the \emph{evolution} of gravitation beyond the CH however is highly non-unique. This corresponds to a severe failure of determinism in General Relativity (GR).

There are different ways to interpret the results of Ref.~\cite{Cardoso:2017soq}. One could take the SCC conjecture in its \emph{conceptual} version, where SCC is purely a mathematical question about General Relativity and its limits. Then the results of Ref.~\cite{Cardoso:2017soq} either signify a failure of SCC, or are superseded by nonlinear effects. 
Here, we have nothing else to add on this purely mathematical question.

Alternatively, one can interpret the SCC conjecture in an \textit{anthropic-astrophysical} sense, where restrictions arising from experimental or observational data (including gravitational waves, BH and cosmological observations, or information arising from particle physics) need to be taken into account. In other words, in such a viewpoint GR would need to be supplemented with all the fields of the Standard Model and perhaps even with quantum-gravity effects. In this context, the following are commonly accepted facts: 

\noindent {\bf i.} first, BHs in our universe are nearly neutral.
Electromagnetic charge is quickly neutralized by either environmental plasma, Schwinger pair-creation or Hawking evaporation~\cite{Cardoso:2016olt}. In light of this, one can question the relevance
of SCC violations in highly charged, non-spinning BH spacetimes;

\noindent {\bf ii.} to {\it form} a charged BH, charged matter is necessary. Thus, SCC violation with only neutral fields is unrealistic, and needs to be generalized to charged fields as well.

The results of Ref.~\cite{Cardoso:2017soq} were followed by various attempts to save the conjecture, supported by observation i.\ or ii.\ above. 
Firstly, it was shown that the Cauchy horizons of rapidly rotating BHs in cosmological backgrounds behave differently from those of highly charged BHs~\cite{Dias:2018ynt}. 
According to Ref.~\cite{Dias:2018ynt}, in Kerr--dS Eq.~\eqref{theoremHintz} remains valid, but now $\beta$ seems to be bounded exactly by $1/2$, with the bound being saturated at extremality. 
Such a result might suffice to save SCC in the context of astrophysical BHs. However, the behavior of rapidly spinning, but weakly charged BHs is unknown, and these may well exist in our universe.

Here, we will discuss another work~\cite{Hod:2018dpx} providing evidence that when point ii.\ above is taken into account and charged scalars are considered, then $\beta<1/2$ in an appropriate region of parameter space, and consequently SCC is upheld. 
This last implication requires, first of all, the validity of~\eqref{theoremHintz} for charged scalars, which does require a justification~\footnote{We stress the fact that the relation~\eqref{theoremHintz} is not universal. In fact, for BHs with vanishing cosmological constant the value of $\beta$ seems to be irrelevant in the context of SCC \cite{LukOhStrongI,LukOhStrongII}.}.
In appendix~\ref{app:beta} we show that Eq.~\eqref{theoremHintz} does generalize in the expected way, with the critical value being, once again, $\beta = 1/2$.
In addition, the methods in Ref.~\cite{Hod:2018dpx} require working in the large-coupling regime $q Q\gg {\rm max}(\mu\,r_+,l+1)$, with $Q$ the BH charge, $q$ the field charge, $\mu$ the scalar field mass, and $r_+$ the radius of the event horizon.

We finish this section by acknowledging yet another interesting recent suggestion to remedy SCC, in the presence of a positive cosmological constant: in Ref.~\cite{Dafermos:2018tha}, it was shown that the pathologies identified in Ref.~\cite{Cardoso:2017soq} become non-generic if one considerably enlarges the allowed set of initial data by weakening their regularity. Although the considered data are compatible with the modern formulation of SCC, we believe that SCC is, in essence, a formation of singularities problem\footnote{In contrast, Weak Cosmic Censorship is concerned with the avoidance of naked singularities.} which is mainly of interest for regular initial data; the mechanism of SCC becomes obscured if one considers initial data which are too ``rough'' (compare with the problem of the formation of shocks in fluid mechanics~\cite{ChristodoulouShocks}).

\noindent{\bf{\em IIa. Charged Scalar Perturbations of RNdS.}}
The purpose of our work is to explore the decay of {\it charged} scalar fields in the full range of charge coupling $q Q$ and various choices of scalar masses $(\mu M)^2$ on those RNdS BH backgrounds which were identified as pathological in~\cite{Cardoso:2017soq}. We will then be addressing concern ii.\ above~\footnote{Note that a deeper understanding of concern ii.\ would also require the study of fermions.}.

We will show that it is not necessary to impose lower bounds on the scalar field mass to obtain $\beta<1/2$.
On the other hand, we will demonstrate that for small charge coupling one can still find regions in parameter space where SCC is violated ($\beta>1/2$).

The background spacetime is a charged RNdS,
\begin{equation}
\label{RNdS_space}
ds^2=-F(r)dt^2+\frac{dr^2}{F(r)}+r^2(d\theta^2+\sin^2\theta d\phi^2)\,,
\end{equation}
where $F(r)=1-{2M}{r^{-1}}+{Q^2}{r^{-2}}-\Lambda r^2/3$. Here, $M,\,Q$ are the BH mass and charge, respectively, and $\Lambda>0$ is the cosmological constant. The surface gravity of each horizon is then
\begin{equation}
\label{surfGrav}
\kappa_\gamma= \frac{1}{2}|F'(r_\gamma)|\;\;,\; \gamma\in\{-,+,c\}\;,
\end{equation}
where $r_-<\,r_+<\, r_c$ are the Cauchy horizon, event horizon and cosmological horizon radius.

A minimally coupled charged massive scalar field on a RNdS background with harmonic time dependence can be expanded in terms of spherical harmonics,
\begin{equation}
\sum_{lm}\frac{\Psi_{l m}(r)}{r}Y_{lm}(\theta,\phi)e^{-i\omega t}\,.
\end{equation}
Dropping the subscripts on the radial functions, they satisfy the equation
\begin{equation}
\label{master_eq_RNdS}
\frac{d^2 \Psi}{d r_*^2}+\left[\left(\omega-\Phi(r)\right)^2-V_l(r)\right]\Psi=0\,,
\end{equation}
where $\Phi(r)=q Q/r$ is the electrostatic potential, $q$ the charge of the scalar field and $dr_*=\frac{dr}{F}$ the tortoise coordinate.
The effective potential for scalar perturbations is
\begin{equation}
\label{RNdS_general potential}
V_l(r)=F(r)\left(\mu^2+\frac{l(l+1)}{r^2}+\frac{F^\prime(r)}{r}\right),
\end{equation}
where $l$ is an angular number, corresponding to the eigenvalue of the spherical harmonics, and $\mu$ the mass of the scalar field.

We are interested in the characteristic quasinormal (QN) frequencies $\omega_{ln}$ of this spacetime, obtained by imposing the boundary conditions~\cite{Berti:2009kk}
\begin{equation}
 \Psi(r\to r_+)\sim e^{- i\left(\omega-\Phi(r_+)\right) r_*}\,\,,\,\,
 \Psi(r \to r_c)\sim e^{ i\left(\omega-\Phi(r_c)\right) r_*}\nonumber \label{bcs}\,.
\end{equation}
The QN frequencies are characterized, for each $l$, by an integer $n\geq 0$ labeling the mode number.

The fundamental mode $n=0$ corresponds, by definition, to the \emph{non-vanishing} frequency with the largest imaginary part and will be denoted by $\omega_0\neq 0$. 

As shown in appendix~\ref{app:beta}, for $q Q\neq0$ the stability of the Cauchy horizon continues to be determined by~\eqref{betaOld}.

We note that the only vanishing mode we find (see results below) corresponds to the trivial mode at $l=0$ and $q=\mu=0$. 
In fact, massless, neutral scalars can always be changed by an additive constant without changing the physics. 
Thus, the zero mode is irrelevant for the question of stability of the CH and consequently must be discarded in the definition of $\beta$  (see the discussion in the end of appendix~\ref{appDefBeta}).

The results shown in the following sections were obtained mostly with the Mathematica package of~\cite{Jansen:2017oag} (based on methods developed in~\cite{Dias:2010eu}), and checked in various cases with a WKB approximation~\cite{Iyer:1986np}.

\noindent{\bf{\em IIb. QNMs of massless, neutral scalar fields.}}
In~\cite{Cardoso:2017soq},
we found 3 qualitatively different families of QNMs: the photon sphere (PS) family, the de Sitter (dS) family and the near extremal (NE) family.
The first two connect smoothly to the modes of asymptotically flat Schwarzschild and of empty dS, respectively, while the last family cannot be found in either of these spacetimes.

Finally, apart from the previous 3 families  (for $q=\mu=0$) there is also  a single orphan mode---the trivial zero mode at $l=0$.

\noindent{\bf{\em IIIa. Charged Massless Scalars.}}
Since the main point of the current work is to investigate if the inclusion of charged matter saves SCC, we will restrict ourselves to choices of near extremal BH parameters identified as problematic in~\cite{Cardoso:2017soq} from the point of view of SCC. 

Since, in~\cite{Cardoso:2017soq}, the dependence on the cosmological constant was found to be minimal (provided that it is positive!), we will restrict  to $\Lambda M^2 = 0.06$ throughout this paper.
We expect our results to be qualitatively independent of this choice.

The BH charges we consider are:
\begin{equation}
\label{EqBHC}
1 - Q/Q_\text{max}  = 10^{-3},\ 10^{-4},\ 10^{-5}\,.
\end{equation}

\begin{figure}[t]
\includegraphics[width=0.48\textwidth]{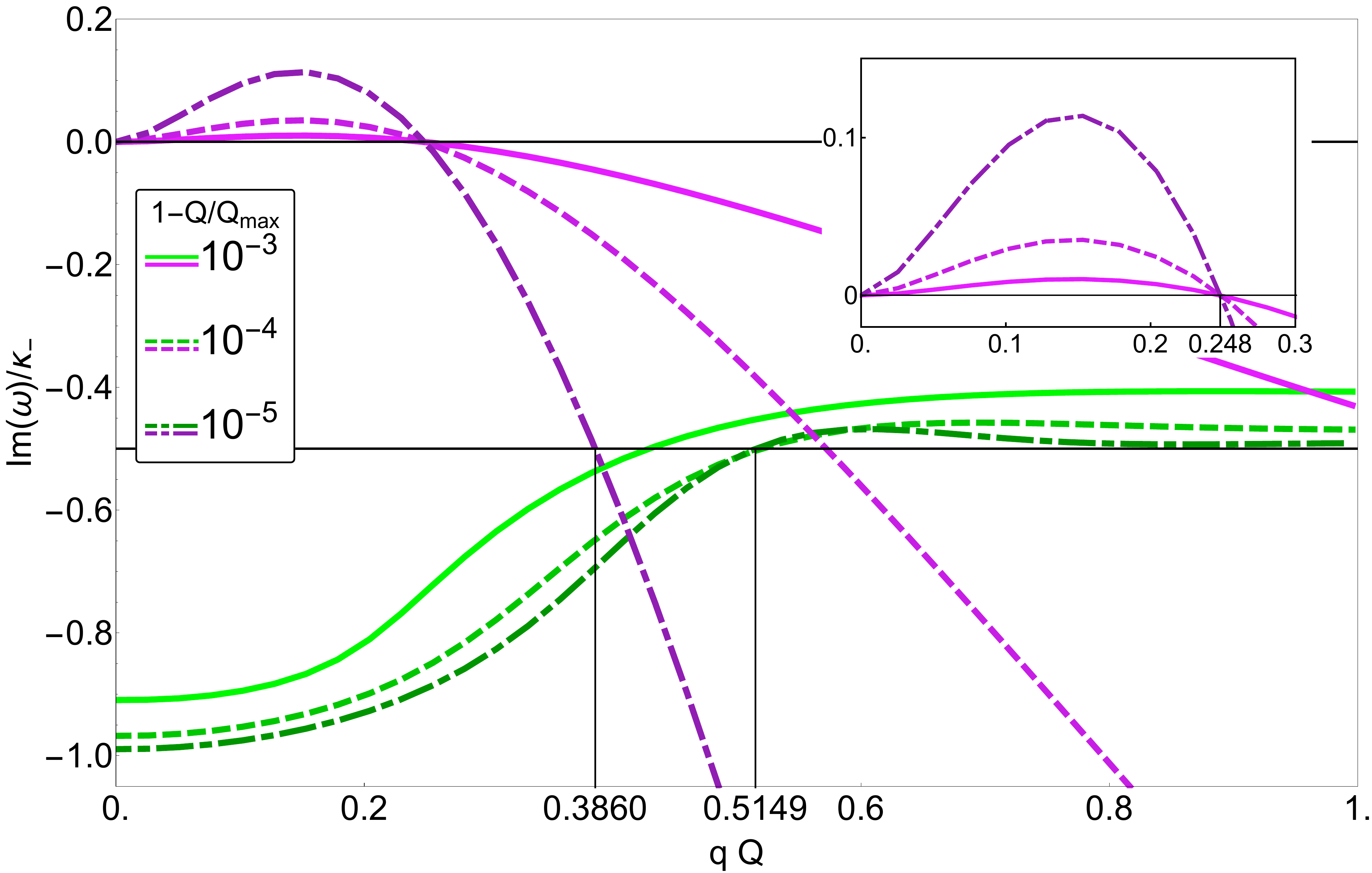}
\caption{The two lowest QNMs of a charged, massless scalar perturbation of a RNdS BH with $\Lambda M^2 = 0.06$ and  $1 - Q/Q_\text{max}  = 10^{-3}, 10^{-4}$, $10^{-5}$, as a function of the scalar charge $q$.
The purple modes originating from $(\omega,q) = (0,0)$ are called superradiant modes, and the green modes are called near extremal modes.}
\label{chargedScalar}
\end{figure}

\begin{figure}[t]
\includegraphics[width=0.48\textwidth]{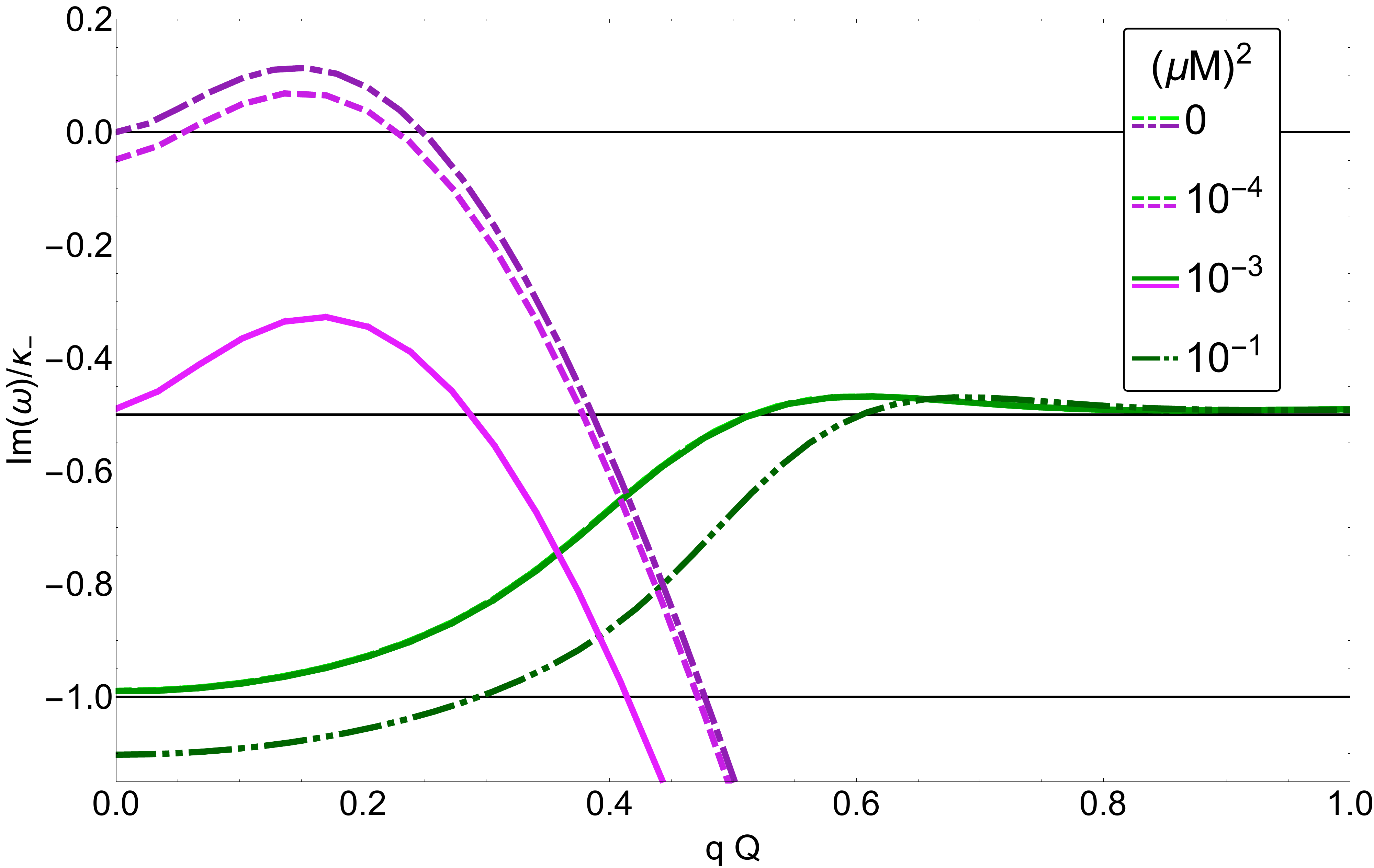}
\caption{Lowest QNMs of a charged scalar perturbation of a RNdS BH with $\Lambda M^2 = 0.06$ and  $1 - Q/Q_\text{max}  =10^{-5}$, as a function of the scalar charge $q$, for various scalar masses $(\mu M)^2 = 0, 10^{-4}, 10^{-3}$ and $10^{-1}$.
The top 3 purple lines are superradiant modes, with the very top one corresponding to the one with the same style in Fig.~\ref{chargedScalar}, and for the largest mass $(\mu M)^2 = 10^{-1}$ the superradiant mode lies outside the plotted range. 
Green lines are near extremal modes, of which the top three overlap.}
\label{chargedScalarSmallMu}
\end{figure}

According to our results (see Fig.~\ref{chargedScalar}, Fig.~\ref{chargedScalarSmallMu} and appendix~\ref{app:largel}), in this parameter range, the dominant mode is a spherically symmetric $l=0$ mode. 
Note that this was already the case for the massless, neutral scalars studied in~\cite{Cardoso:2017soq}.

Note also that, in view of our focus on near extremal charges, the PS and dS families are considerably subdominant and, consequently, are not seen here.  

Consider first charged, {\it massless} scalars. In Fig.~\ref{chargedScalar}, the two most dominant QNMs are shown, for each of the BH charges in~\eqref{EqBHC}.
The green lines correspond to NE modes, $\omega_{\rm NE}$~\cite{Cardoso:2017soq}.
At $q=0$, we find $1/2<\beta<1$, in agreement with our previous results~\cite{Cardoso:2017soq}. For $q>0$, the NE modes are initially subdominant but eventually, for sufficiently large $q Q$, become dominant and such that $\text{Im}(\omega_{\rm NE})/\kappa_->-1/2$, so $\beta<1/2$ for such $q Q$.
This corroborates the arguments of~\cite{Hod:2018dpx} and extends them to the massless scalar setting. In particular, we conclude that the large scalar mass condition of~\cite{Hod:2018dpx} is not necessary to guarantee that $\beta<1/2$: large $q Q$ suffices.

The purple lines complicate the story. We call the corresponding QNM superradiant (SR) modes, as they are associated, for small $q Q$, with a superradiant instability~\cite{Brito:2015oca} 
(for larger $q Q$ they are decaying modes). These modes were seen for the first time in Ref.~\cite{Zhu:2014sya} and further analyzed in Ref.~\cite{Konoplya:2014lha}. 
They originate from the trivial mode of the massless, neutral scalar, at $l=0$, which corresponds to nothing more than a constant shift in the scalar field. 
When we add charge or mass, the corresponding wave equation no longer admits constant solutions, the trivial mode disappears and gives rise to the dynamical mode seen in Fig.~\ref{chargedScalar}. 
For small coupling $q Q$, the SR modes are unstable, $\text{Im}(\omega)>0$, with the maximum imaginary part increasing with the size of the BH charge. 
This linear instability suggests that under evolution by the full Einstein equations, coupled with the fields under consideration here, even the \emph{exterior} of our RNdS BH will be severely unstable; thus, we cannot infer anything about SCC in this case.

However, it is also apparent from Fig.~\ref{chargedScalar} that the SR modes cross the $\text{Im}(\omega) = 0$ at $q Q \approx 0.248$ (which to a good approximation is independent of
the particular BH charge). The modes then become stable and, eventually, subdominant---the dominant mode becomes the one arising from the NE family.
Very interestingly however, by inspection of Fig.~\ref{chargedScalar}, we find that there are choices of parameters for which $\beta>1/2$: this happens for instance when $1 - Q/Q_\text{max}  = 10^{-5}$ and $0.386 < q Q < 0.515$. 
We remark that for massless scalars, we once again find that $\beta$ is bounded, never reaching unity.

\noindent{\bf{\em IIIb. Charged Massive Scalars.}}
%
We now focus on a BH charge satisfying $1 - Q/Q_\text{max}  = 10^{-5}$, and study the effect of adding a scalar mass to the QNM landscape. 
We present part of our results in Fig.~\ref{chargedScalarSmallMu} by showing the two most dominant QNMs for a selection of scalar masses.
The effect of the mass is to decrease the imaginary part of both modes.
This means that massive scalars decay faster; consequently, the larger the scalar mass, the harder it becomes for its fluctuations to restore SCC. 

The strongest effect here is in the SR modes. These are highly sensitive to the mass, which moves the imaginary part downwards by an approximately constant shift.
(In the uncharged case $q Q=0$, this was justified rigorously for small $\mu>0$ in \cite{HintzVasySemilinear}.)
As a consequence, for non-zero mass, the superradiant instability is no longer present for sufficiently small charges $q$, but resurfaces when $q$ exceeds some positive lower bound. 
We note that the existence of unstable SR modes with non-vanishing scalar mass (which follows from their existence in the massless case by the continuous dependence of the QNM spectrum on all parameters) is, to the best of our knowledge, being numerically detected here for the first time. These modes are easy to miss in view of their large sensitivity to changes in the mass. In fact, for larger, but still very small, masses, the superradiant instability is no longer present. 

The NE mode is much less sensitive to the mass (the first 3 green lines in Fig.~\ref{chargedScalarSmallMu} lie on top of each other).
While it also moves down, it continues to be the case that for large enough $q Q$, of order 1, and the mass range considered, $\beta<1/2$.
The limiting value of $\beta$ at large $q Q$ seems to be independent of the mass.

Finally, notice that for the largest mass presented, $(\mu M)^2 = 10^{-1}$, the SR mode is outside of the plotted range, and the NE mode is below $\text{Im}(\omega/\kappa_-) = -1$, indicating a $\beta > 1$, found here for the first time.
Note that, although it seems that the NE mode might continue increasing its negative imaginary part with increasing scalar mass, we do expect $\beta$ to remain bounded, since at some point the photon sphere mode will become dominant, 
and this is independent of the mass at large $l$. 
Furthermore, In empty dS, $\beta$ remains bounded when $\Lambda\mu^2\to\infty$, as follows from~\cite{VasyWaveOndS}.

\noindent{\bf{\em IV. Conclusions.}}
In a recent paper~\cite{Cardoso:2017soq} we presented evidence for the failure of SCC for highly charged RNdS BHs under neutral scalar perturbations.  This was achieved by relying on Eq.~\eqref{theoremHintz} and performing a thorough numerical computation of $\beta$. Here, following a suggestion in Ref.~\cite{Hod:2018dpx}, we extend our analysis to charged (massless and massive) scalars.  

To obtain a quantitative formulation of (a linearized version of)  SCC we started by showing that, by suitably extending the definition of $\beta$, Eq.~\eqref{theoremHintz} remains valid for charged and massive scalar perturbations. 
We then performed a detailed numerical computation of the dominant QNMs in RNdS, for choices of BH parameters identified as problematic in Ref.~\cite{Cardoso:2017soq}, while taking into account the entire range of coupling constants $q Q\geq 0$ and several choices of scalar masses. 
From this we can then compute $\beta$ and infer about SCC, at least in the cases where we have mode stability $\text{Im}(\omega_0)<0$.    

Our main results are plotted in Figs.~\ref{chargedScalar} and~\ref{chargedScalarSmallMu}, and our conclusions can be summarized as follows: 
\begin{enumerate}
\item For all choices of scalar mass and large enough charge coupling we get $\beta<1/2$. Consequently, our linear analysis suggests that SCC should be valid, in the corresponding parameter ranges. This is in line with Ref.~\cite{Hod:2018dpx}, but here we show that the result also holds for small masses.

Superimposing neutral and charged scalar perturbations, the smaller of the two values of $\beta$ for the two types of perturbations (namely $\beta_{|q=0}>1/2$ and $\beta_{|q\gg 1}<1/2$) is the one relevant for SCC. Consequently the expected failure of SCC for uncharged scalars is put into question as soon as we add a charged scalar field with sufficiently strong coupling.

\item Nonetheless, for all choices of scalar masses we always find an interval of coupling charges for which $\beta>1/2$, which predicts a potential failure of SCC in this setting. 

Moreover, even if we add neutral perturbations we will get $\beta_{|q=0}>1/2$ and $\beta_{|q \neq0}>1/2$ and the situation remains alarming for SCC.  

\item Finally, for large scalar masses and small charges we get, for the first time, $\beta>1$. Recall that this is related to bounded curvature and therefore opens the possibility to the existence of solutions to the Einstein--Maxwell--KG system  with a scalar field satisfying Price's law and bounded curvature across the Cauchy horizon.~\footnote{Spherically symmetric solutions of the Einstein-Maxwell scalar system with bounded curvature where constructed in Ref.~\cite{Costa:2014aia}, but these have a compactly supported scalar field along the event horizon.} 

Nonetheless, this should be a non-generic feature: if we once again superimpose a neutral scalar perturbation---as a proxy for a linearized gravitational perturbation---we will get $\beta_{|q=0}<1$, which should be enough to guarantee the blow-up of curvature.

\end{enumerate}

We end with some final comments.

First, the charged matter could just as well be fermionic instead of scalar. Note that fermions do not have a superradiant instability, so the entire range of fermion charge parameters is now open for the study of SCC at this linear level. In particular, it  would be interesting to see if charged fermions also have the potential to restore SCC at large charge. We plan to pursue this study in the near future. 

Second, during the last stages of this work we were informed by the authors of Ref.~\cite{Hongbao} that the superradiant mode, unlike the near extremal, is sensitive to the cosmological constant: 
for large enough cosmological constant it appears to be absent. 
A quick check indicates that our chosen value of $\Lambda M^2 = 0.06$ is close to the value where the superradiant mode is the most unstable.
\footnote{For $1 - Q/Q_\text{max} = q Q = 10^{-2}$ the massless superradiant mode has $\text{Im}(\omega)/\kappa_- = 2.75 \times 10^{-5}$ when $\Lambda M^2 = 0.06$, and obtains its maximum of $\text{Im}(\omega)/\kappa_- = 3.07 \times 10^{-5}$ when $\Lambda M^2 = 0.0426$.}
Hence for different $\Lambda M^2$ we expect the role of this mode to be either similar or smaller (for $\Lambda = 0$ the instability is absent~\cite{Konoplya:2013rxa}), and thus for it to be equally hard or easier to find a regime where $\beta > 1/2$.

Third, one might also argue, as is done in \cite{Hod:2018dpx}, that for physical black holes made from charged matter coming from the standard model we must have $q Q \gg 1$; hence SCC is, according to the presented results, expected to be satisfied. We stress that even with this input we remain in the realm  of the \emph{conceptual} version of SCC (as described in the introduction) since the input is only relevant when the conjecture is in danger and this only happens for  highly charged BHs.  

Finally, in the final stage of preparation \cite{Dias:2018etb} appeared, where the gravitational and electromagnetic perturbations are analysed and found not to save SCC.
In fact for these perturbations it is found that $\beta$ can exceed not just $1/2$ but even 1 and 2, making the charged matter studied here the dominant mode.
 
Regardless of the approach to SCC that the reader subscribes to, the results presented here indicate at least a growing level of sophistication required for the Cosmic Censor, and the situation regarding Strong Cosmic Censorship, in the presence of a positive cosmological constant, remains subtle!

\textbf{Note added:}
Shortly after submitting our results to arXiv, a new paper~\cite{Dias:2018ufh} appeared claiming that even at large scalar charge, there are wiggles in the dominant QNM, as a function of scalar and/or black hole charge, that cause further regions in parameter space where SCC is violated.
We do not see such a phenomenon for our value of the cosmological constant, $\Lambda M^2 = 0.06$, but our numerics cannot conclusively rule it out. We do note that~\cite{Dias:2018ufh} has only investigated large $\Lambda$ (in particular $\Lambda M^2 > 1/9$, for which RNdS black holes have a minimum non-vanishing charge), so an absence of these wiggles at our $\Lambda M^2 = 0.06$ would not be inconsistent.

\noindent{\bf{\em Acknowledgments.}}
The authors acknowledge financial support provided under the European Union's H2020 ERC 
Consolidator Grant ``Matter and strong-field gravity: New frontiers in Einstein's theory'' grant 
agreement no. MaGRaTh--646597. 
This project has received funding from the European Union's Horizon 2020 research and innovation programme under the Marie Sklodowska-Curie grant agreement No 690904.
The authors would like to acknowledge networking support by the GWverse COST Action CA16104, ``Black holes, gravitational waves and fundamental physics.''
J.L.C.\ acknowledges financial support provided by FCT/Portugal through UID/MAT/04459/2013 and grant (GPSEinstein) PTDC/MAT-ANA/1275/2014.
Part of this research was conducted during the time P.H.\ served as a Clay Research Fellow.

\begin{appendix}

\section{The definition of $\beta$ for charged scalars.}\label{app:beta}
\label{appDefBeta}
Since we are mainly interested in the problem of finding regimes where SCC is potentially \emph{violated}, and study this by means of QNM calculations, we need two ingredients: 1) `resonance expansions', i.e.\ expansions of linear waves into mode solutions of the underlying wave equation, and 2) the relationship between QNMs and a lower bound on the regularity of the corresponding mode solutions at the CH.

As for 1), the validity of full resonance expansions has only been established for scalar waves on Schwarzschild--de~Sitter and Kerr--de~Sitter spacetimes \cite{BonyHaefner,Dyatlov:2011jd}; for our purposes, resonance expansions with remainder terms decaying at an exponential rate given by a WKB approximation are sufficient, as the dominant modes have much slower decay rates (as discussed above). For $\Lambda>0$, such partial resonance expansions, and the discreteness of the QNM spectrum, hold under very general conditions~\cite{VasyMicro,DyatlovSpectralGaps}: non-zero surface gravities of the horizons of the BH exterior (i.e.\ event and cosmological horizon), and a condition on the skew-adjoint part of the relvant wave operator at the trapped set (photon sphere) which is verified for a large class of tensorial wave equations on BH backgrounds~\cite{HintzPsdoInner,Hintz:2016KNdS}.

As for 2), we recall that Eq.~\eqref{master_eq_RNdS} arises from the spherical harmonic decomposition of
\[
  P(\Psi/r) = 0,\ \ 
  P=(\nabla^\nu-i q A^\nu)(\nabla_\nu-i q A_\nu)-\mu^2,
\]
where $A=-(Q/r)d t$ is the vector potential. To determine the regularity of mode solutions up to the CH, we first need to use the $U(1)$ gauge freedom to transform $P$ into an operator with \emph{smooth} coefficients up to the CH---note that $A$ becomes singular at the horizons: conjugating $P$ by $e^{i q\chi}$ for a real-valued function $\chi$ amounts to replacing $A$ by $\tilde A=A+d\chi$. Choose $\chi=\chi(r)$ so that $\tilde A=-(Q/r)d t_*$, where $t_*=t-G(r)$, $G'(r)=1/F(r)$ is smooth across CH and has past causal differential near the CH. Extending the definition of $t_*$ suitably to the region $r_-\leq r\leq r_c$, the level sets $t_*=c$ tend to future timelike infinity as $c\to\infty$; thus, QNMs are those $\omega\in\mathbb C$ for which there exists a mode solution $e^{-i\omega t_*}X(r,\theta,\phi)$,
\begin{equation}
\label{EqMode}
  \tilde P(e^{-i\omega t_*}X)=0,\ \ 
  \tilde P=(\nabla^\nu-i q \tilde A^\nu)(\nabla_\nu-i q \tilde A_\nu)-\mu^2,
\end{equation}
which is smooth across $r=r_+$ (event horizon) and $r=r_c$ (cosmological horizon). Our task is to determine the regularity of $X$ at the CH. By spherical symmetry (or much more robust arguments, only relying on the non-zero surface gravity of the CH \cite{Hintz:2015jkj}), one can show that $X$ is \emph{conormal} to the CH, which is to say that $X$ and arbitrarily many derivatives of $X$ along vector fields \emph{tangent} to the CH have a bound $\lesssim|r-r_-|^{-C}$ with $C$ fixed. Thus, the main terms of $\tilde P$ are those involving \emph{radial} derivatives (which are transversal to the CH). Keeping only these terms and freezing coefficients at $r=r_-$, one gets a regular-singular ODE in $r$:
\begin{align*}
  0 &= e^{i q\chi}\Bigl( (F\partial_r)^2+2 i\omega F\partial_r - \frac{2 q Q\omega}{r_-} + \frac{q^2 Q^2}{r_-^2}\Bigr) e^{-i q\chi} X \\
    &= 4\kappa_-^2 (F\partial_F)\biggl(F\partial_F - \Bigl(\frac{i\omega}{\kappa_-}-\frac{i q Q}{\kappa_- r_-}\Bigr)\biggr)X + \text{l.o.t};
\end{align*}
in the second line, we switched to $F$ as a radial coordinate. Therefore, we can significantly sharpen our grip on $X$: it is in general the sum of two terms, one with constant, or smooth, asymptotics at the CH (since $F\partial_F(1)=0$), and another one with asymptotics.
\begin{equation}
\label{EqSing}
  |F|^{\frac{i\omega}{\kappa_-}-\frac{i q Q}{\kappa_- r_-}}.
\end{equation}
(There is an additional factor of $\log|F|$ when the exponent is an integer.) At the CH, where $F=0$, this lies in the Sobolev space $H^{1/2+\beta(\omega)-\epsilon}$ for all $\epsilon>0$, where
\[
  \beta(\omega) = \kappa_-^{-1}\text{Im}(-\omega+q Q/r_-) = -\text{Im}(\omega)/\kappa_-.
\]
The contribution to $X$ of a term with asymptotics~\eqref{EqSing} may well vanish for certain values of $\omega$, in which case $X$ is in fact smooth! This happens in the case $q Q=0$ and $\omega=0$, when $X\equiv 1$ solves~\eqref{EqMode}, which is why this zero mode was excluded from the definition of $\beta$ in the uncharged case. (It is not known whether this is the only special situation in which $X$ is smooth at the CH.) On the other hand, setting $\beta=\inf_\omega\beta(\omega)$ (over all QNMs $\omega$ when $q Q\neq 0$, or excluding $\omega=0$ when $q Q=0$), we definitely know that all modes, and thus the solution of the wave equation itself, have regularity $H^{1/2+\beta-\epsilon}$ at the CH. This provides the justification for our search for BH parameters for which $\beta>1/2$.

\section{Higher $l$ modes}\label{app:largel}
\begin{figure}[t]
\includegraphics[width=0.48\textwidth]{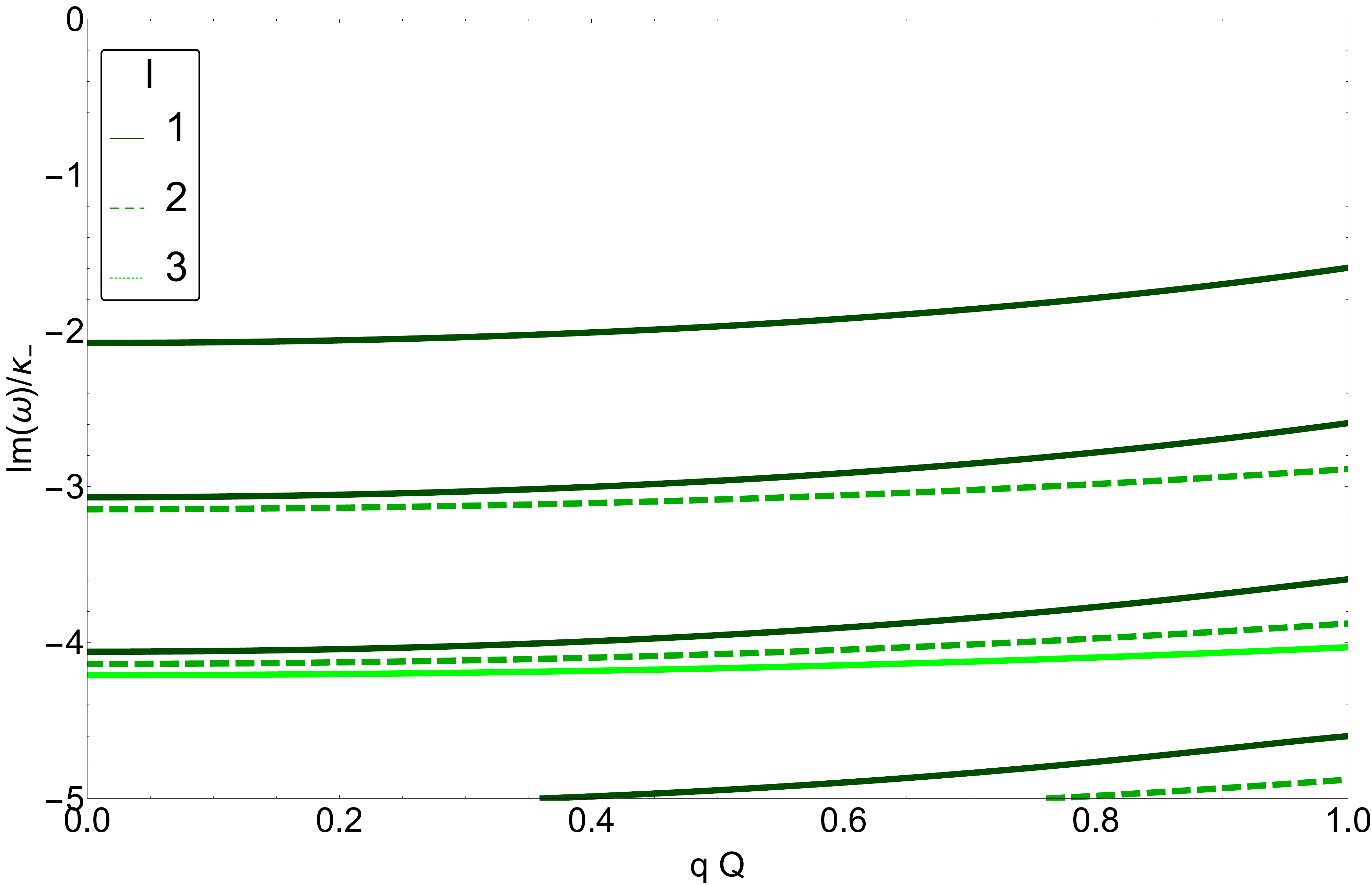}
\caption{Higher $l$ QNMs of a charged, massless scalar perturbation of a RNdS BH with $\Lambda M^2 = 0.06$ and  $1 - Q/Q_\text{max}  = 10^{-5}$, as a function of the scalar charge $q$.}
\label{higherlmassless}
\end{figure}
In this appendix we verify the expectation that the higher $l$ QNMs do not affect Strong Cosmic Censorship. 
In Fig.~\ref{higherlmassless} we show the $l=1,2,3$ modes. 
These are all modes of the near extremal family, since they are present already at $q=0$ and they follow their pattern. No modes of the other families are present at this range.
The dependence on $q$ is rather mild, and in particular they do not come near the dominant $l=0$ mode.

While Fig.~\ref{higherlmassless} shows the modes for massless scalars, we have done the same check for the massive. Of the masses considered, the modes for $(\mu M)^2 = 10^{-4}$ and $10^{-3}$ are visually indistinguishable from the massless ones, and for $(\mu M)^2 = 10^{-1}$ they lie just below the ones presented.

Finally, one might worry if for even larger $l$ the photon sphere modes will become dominant.
To address this we have computed by WKB approximation, which is expected to become very accurate in the large $l$ limit, the modes at $l=100$. 
The dominant mode that we find for the parameters of Fig.~\ref{higherlmassless} and $ q Q = 0.45$ is $\omega / \kappa_- = 5472 - 18.35 i$, which we confirmed using \cite{Jansen:2017oag}.
Furthermore, we have checked all other values of scalar charge and mass considered and found this value to be largely independent of those parameters.

This is very far from the $l=0$ mode, so we are convinced that throughout the parameter space considered, $l=0$ indeed gives the dominant mode.

\end{appendix}

\bibliography{references,referencesNeutral}

\end{document}